\documentclass[journal=langd5,manuscript=article]{achemso}
  \usepackage[version=4]{mhchem}
 \usepackage{chemformula}
 \usepackage{float}
 \usepackage[labelfont=bf,labelsep=period]{caption}
 \usepackage{sectsty}
 \usepackage[utf8]{inputenc}
 \usepackage{array}
 \newcolumntype{P}[1]{>{\centering\arraybackslash}p{#1}}
 \usepackage{geometry}
 \usepackage{setspace}
 \usepackage{xkeyval}
 \usepackage{xcolor}
 \usepackage{natbib}
 \usepackage{multirow}
 \usepackage{tablefootnote}
 \usepackage[section]{placeins}
\usepackage{lineno,hyperref}
\modulolinenumbers[5]
\DeclareUnicodeCharacter{2212}{-}
\author{Zixiang Wei}
\affiliation{Department of Chemical and Materials Engineering, University of Alberta, Alberta T6G 1H9, Canada}
\email{zwei3@ualberta.ca}
\author{Jae Bem You}
\affiliation{Department of Chemical and Materials Engineering, University of Alberta, Alberta T6G 1H9, Canada}
\author{Hongbo Zeng}
\affiliation{Department of Chemical and Materials Engineering, University of Alberta, Alberta T6G 1H9, Canada}
\author{Xuehua Zhang}
\affiliation{Department of Chemical and Materials Engineering, University of Alberta, Alberta T6G 1H9, Canada}

\title {Interfacial Partitioning Enhances Microextraction by Multicomponent Nanodroplets}

\keywords{surface nanodroplets, liquid-liquid extraction, partition coefficient, colorimetric reaction, interfaces}

\begin{document}
\begin{abstract}

The sensitive and reliable in-droplet chemical analysis benefits from the enhanced partition of an analyte into the droplets. This work, we will show that chemical reactions in surface nanodroplets can shift the partition of analytes from a highly diluted solution to the droplets. Seven types of organic acids with partition coefficients (LgP) ranging from -0.7 to 1.87 are used as model analytes dissolved in an oil solution that are extracted from the flow into aqueous nanodroplets immobilized on a substrate. The timescale of integrated extraction and reaction in droplets was represented by the decoloration time of the droplets. Our results show that the effective distribution coefficient of the analyte can be decreased by 3 to 11 times of the distribution coefficient of the analyte in the bulk liquids. The principle behind the significantly shifted partition is proposed to be enhanced the transfer of the analyte across the droplet surface. The chemical reaction in the droplets enhances the partition of the analyte from a highly diluted solution. Our results show that the interfacial behavior of the analyte may be advantageous as it may improve extraction and partition. Such enhanced extraction may be leveraged for sensitive chemical detection using reactive droplets.

\end{abstract}

\section{Introduction}

Sample pretreatment plays a vital role in chemical identification and quantification with reliability and sensitivity.  
\cite{rezaee2006dllme,yan2013dllme,alexovivc2017automation} Many analytic techniques require the analytes to be separated from interfering substances in the sample or be transferred to a suitable medium for meaningful and high-quality measurements.  Standard bulk protocols for separation of the analyte from a complex samples consume a large amount of toxic solvents and therefore are regarded as one of the most polluting steps in chemical analysis. 
It is estimated that the solvent waste from liquid chromatography alone is about 34 million
liters in a year.\cite{miller2012preparative} \\
Integrating separation and detection of the analyst in small droplets is a way to minimizing solvent use, automating and streamlining sample pretreatment and detection, and developing greener and cleaner analytic protocols. Transferring the analyst to small droplets can be fast and efficient across the interface of two immiscible phases due to the large surface area to volume ratio of the droplets. Moreover, the droplets may provide a discrete local environment where indicative reactions with the extracted analyte may take place to produce detectable signatures for in-droplet analysis. However, a fundamental question to be addressed for in-droplet chemical analysis is how the chemical reaction in droplets impacts the dynamics and the equilibrium of the analyst partition from the surrounding medium.\\
Liquid-liquid microextraction is based on the solubility difference to extract the compunds from the surrounding phase into microdroplets.\cite{comer2001lipophilicity} The concentration of the compound in  microdroplet is related to the droplet size, solubility of the compound and the extraction time.\cite{jeannot1996solvent,wei2020integrated} Assuming the droplet is water and the compound is dissolved in oil at the concentration of $C_o$, $C_{w(t)}$ is the concentration of the compound in the water droplet at time $t$.
\begin{equation}
\frac{dC_w(t)}{dt}=\frac{A}{V_w}{\beta}(\frac{C_{o}}{p}-C_w(t))
\label{e0}
\end{equation}
Here $V_w$ is the volume of the droplet and $A$ is the interface area where the partition of the compound takes place.   $\beta$ is a coefficient that describes the interfacial transfer of the compound from the surrounding solution into the droplet.  At thermodynamic equilibrium, the enrichment efficiency is limited by the partition coefficient $p$ ($\frac{C_{o}}{C_{w(f)}}$), defined as the ratio of the analyte solubility in the extractant liquid and the aqueous solution. Eq. \ref{e0} suggests that for a given concentration of the surrounding solution $C_{o}$, the concentration of the solute in extractant droplets increases faster for smaller droplets. \\

As separation and extraction of the analyte is coupled with chemical reaction, the rate of the entire process is determined by the slowest rate-limiting step. The dimensionless number Damköhler number (Da) relates the timescale of chemical reaction to that of the mass transport in a system. \cite{fogler2010essentials} However, the reaction rate in droplets is a complicated topic that has not been fully addressed in the literature, even for some basic and well-known chemical reactions. \\
Many chemical reactions are reported to be usually much faster in smaller droplet than in the bulk \cite{li2020speeding,dyett2020accelerated}, such as acid-base reaction,\cite{li2020speeding} dehydration,\cite{li2018accelerated} organometallic reactions\cite{fedick2019screening}, and among others. The reaction rate in microdroplets may be accelerated to a factor of one million, \cite{wei2020accelerated,lee2015microdroplet} while other reactions that requires catalysts in the bulk may take place spontaneously in micro-sized droplets without a catalyst. \cite{nam2017abiotic}  Accelerated chemical reactions in droplets are also confirmed under various configurations,\cite{li2018accelerated,girod2011accelerated,bain2016accelerated,bain2015accelerated} including spray droplets\cite{girod2011accelerated,fenn1990electrospray}, colliding droplet \cite{lee2015microdroplet,lee2017microdroplet}, or levitated droplets, \cite{li2018accelerated, crawford2016real} and among others. \cite{chen2008organic,chen2006cis}\\
To date, the mechanism that leads to the accelerated chemical reaction in microdroplets remains elusive.\cite{wei2020accelerated}. The hypothesis is that the molecules are more energetic at the droplet surface, \cite{zhou2018fluorescence} analogous to the enhanced rate of reaction and product collection at the liquid-liquid or liquid-gas interfaces \cite{wei2020accelerated}. Perhaps the interface region of the droplets may also have a faster refresh rate than in bulk liquid. \cite{fallah2014enhanced} Apart from the chemical reaction rate in the droplets, chemical species at the interface may also have a profound influence on the thermodynamic equilibrium of the partition of the compound in the droplets \cite{fallah2014enhanced}, especially for those amphiphilic molecules with a strong affinity to the interface.  \cite{rezaee2010evolution,wells2020high}. \\


Recently surface nanodroplets have emerged to be a well-controlled platform for integrated droplet extraction and chemical analysis. These nanodroplets are ultrasmall droplets on substrates submerged in an immersible liquid surrounding \cite{zhang2015formation}. Here nanoscale refers to the maximal height of droplets that are tens to several hundreds of nanometers, although the base diameter of the droplets ranges from several hundred nanometers to tens of micrometers. \cite{li2019functional}  Several characteristics of surface nanodroplets are advantageous for integrated extraction and analysis. With such small dimensions surface nanodroplets remain stationary for long time under a flow, complementary to extracting microdroplets that flow in microfluidic devices. The concentration of the analyte in the surrounding that surface nanodroplets are exposed to can be easily mediated by the flow condition of the analyte solution. Furthermore, the concentration of the reactive components in surface nanodroplets can be tuned during a bottom-up nanodroplet production by the solvent exchange method. \cite{qian2019surface,li2018formation}  \\
In our latest work \cite{wei2020integrated}, surface nanodroplets of an aqueous solution were applied to separate and quantify mixtures of organic acids from oil solutions. The originally blue droplets lost color due to the reaction between the extracted acid and the halochromic compound in the droplets. Interestingly, the timescale of the reaction as reflected by decoloration time was found to be specific to the acid type as well as the combination of acids in a mixture of the same pH level. Although the acid specificity of droplet decoloration time may be possibly related to the interfacial properties of individual acids,  it remains unclear how interfacial phenomena are intertwined with microextraction and the reaction of the analyte in the droplets. \\

In this work, we try to address the role of interfacial partitioning in integrated extraction and chemical reaction in colorimetric detection by multicomponent nanodroplets. Our model compounds are organic acids that are extracted into surface nanodroplets and react with a pH indicator dissolved in the droplets.  The timescale of the colorimetric reaction in the nanodroplets will be examined as the droplets are exposed to diluted and concentrated solutions of seven types of acids with a wide range of partition and solubility properties. Significant shifting in acid partitioning is revealed at low concentrations of acids in the surrounding solution. A unified theoretic model is proposed to rationalize the rate of integrated extraction and reaction in droplets and the enhanced partition of the acids. This work may contribute to better understanding of extraction efficiency of reacting droplets, in particular, on the coupling  between interfacial transfer and chemical reaction with highly diluted analytes. The insights from this work may be valuable for downscaling the processes of separation in sensitive detection by using nanodroplets. 


\section{Experimental section}

\subsection{Chemicals and materials}  
Two groups of acids were tested for nanoextraction and colorimetric reaction. Group one consisted of alkyl acids, including acetic acid (99.7\%), n-butyric acid (99.0\%), valeric acid (98.0\%) and heptanoic acid (98.0\%). Group two consisted of aromatic acids, including gallic acid (97.5\%), 4-hydroxybenzoic acid (PHBA, 99\%), and benzoic acid (99.5\%). Acetic acid and heptanoic acid were purchased from Fisher Scientific (Canada). Valeric acid, n-butyric acid, gallic acid, PHBA and benzoic acid were purchased from Sigma-Aldrich (Canada). Some relevant properties of these organic acids (dissociation constant (pKa) and partition coefficient(LgP) along with their chemical structures are listed in Table \ref{t1}. When LgP \textgreater 0, the acid is unfavorable to the water, while LgP \textless 0 is favorable to the water.\\ 
1-Octanol (95 $\%$, Fisher scientific, Canada) was used as the solvent for droplet formation and for the analyte solution. Ethanol (90 $\%$, Fisher scientific, Canada) was used as a co-solvent in our ternary liquid system (1-octanol, water and ethanol). The halochromic compound (pH indicator) was Bromocresol green (BG), which changes color from blue to yellow at a pH of 5.4 to 3.8. Water was obtained from the Milli-Q water purification unit (Millipore Corporation, Boston, MA, USA).  Silicon wafer (University Wafer, USA) coated by APTES (3-aminopropyl triethoxysilane) was cut to 1.3 cm $\times$ 1.5 cm as the substrate in our fluid chamber. The procedure for preparing APTES coating was followed identically to the reported work \cite{APTES1}. A frame of silicon rubber was used as a spacer to control the channel height of the fluid chamber to 300 $\mu$m. A transparent glass slide was cleaned by piranha solution (30\% $H_2O_2$ and 70\% $H_2SO_4$)(caution: piranha solution is highly caustic!) and then used to seal the fluid chamber as cover.
\begin{table}[h]
  \centering
  \caption{Dissociation constant (pKa), Partition coefficient (LgP), critical concentration ($C^*_w$) and molecule structure (MS) of the acids used in this work.}
  \begin{tabular}{p{1cm}|p{1.5cm}|p{1.5cm}|p{1.5cm}|p{1.7cm}|p{1.5cm}|p{1.5cm}|p{1.5cm}}
    \hline
    Acids & Acetic acid & N-butyric acid & Valeic acid& Heptanoic acid& Gallic acid& PHBA& Benzoic acid\\ \hline
    Pka&4.76&4.82&4.84&4.4&4.4&4.54&4.20\\
    \hline
    LgP&-0.7&0.79&1.39&2.42&0.7&1.58&1.87\\
 \hline
    MS&\begin{minipage}[b]{0.09\columnwidth}
		\centering
		\raisebox{-.6\height}{\includegraphics[width=\linewidth]{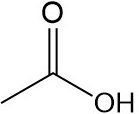}}
	\end{minipage}&\begin{minipage}[b]{0.10\columnwidth}
		\centering
		\raisebox{-.6\height}{\includegraphics[width=\linewidth]{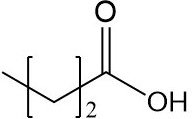}}
	\end{minipage}&\begin{minipage}[b]{0.1\columnwidth}
		\centering
		\raisebox{-.6\height}{\includegraphics[width=\linewidth]{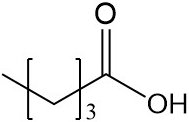}}
	\end{minipage}&\begin{minipage}[b]{0.1\columnwidth}
		\centering
		\raisebox{-.6\height}{\includegraphics[width=\linewidth]{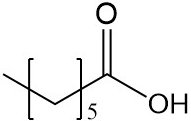}}
	\end{minipage}&\begin{minipage}[b]{0.09\columnwidth}
		\centering
		\raisebox{-.6\height}{\includegraphics[width=\linewidth]{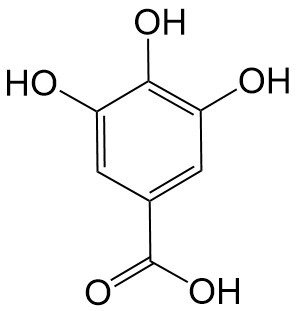}}
	\end{minipage}&\begin{minipage}[b]{0.045\columnwidth}
		\centering
		\raisebox{-.6\height}{\includegraphics[width=\linewidth]{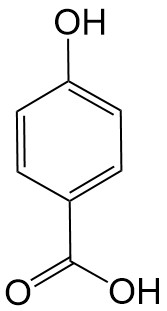}}
	\end{minipage}&\begin{minipage}[b]{0.045\columnwidth}
		\centering
		\raisebox{-.6\height}{\includegraphics[width=\linewidth]{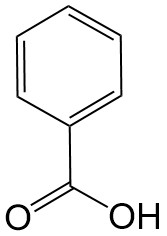}}
	\end{minipage}
\\\hline
  \end{tabular}
  \label{t1}
\end{table}

\subsection{Formation of functional surface nanodroplets}
\begin{figure}[htp]
\centering
	\includegraphics[trim={0cm 0cm 0cm 0cm}, clip, width=0.75\columnwidth]{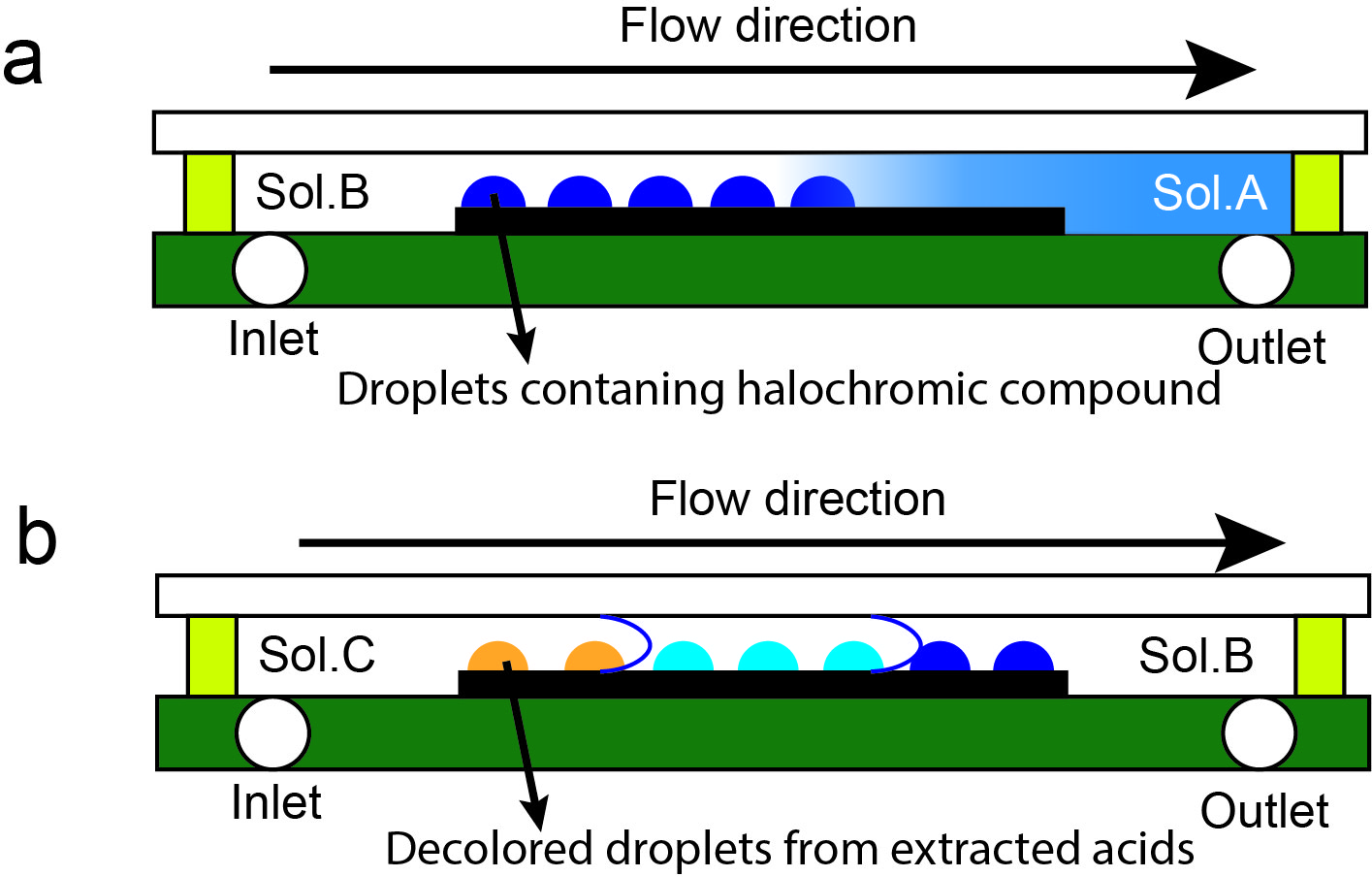}
	\caption{ Sketch of forming droplets (a) and the acid extraction and the colorimetirc reaction in the droplets (b).}
	\label{fx0}
\end{figure}

Following the process illustrated in Figure \ref{fx0}a, aqueous droplets containing BG are formed on a substrate in a microfluid chamber by solvent exchange. The protocol is the same as our previous report.\cite{wei2020integrated}. Colored surface nanodroplets were preformed in a microchamber by the solvent exchange\cite{li2019functional,wei2020integrated}, as sketched in Figure \ref{fx0} (a). In the process of the solvent exchange, solution A was firstly injected into the fluid chamber to completely fill it. Due to the solubility difference of the 1-octanol in two solvents, the colored surface droplets were then formed during the process of replacing of solution A with solution B at a constant flow rate of 15 ml/hr. Solution A for solvent exchange was prepared by mixing 1-octanol, ethanol, and water at a volume ratio of 60:26:14. Then BG was added into solution A at the concentration of 0.002 g/ml. The other two concentrations of BG in solution A, 0.02 g/ml and 0.0004 g/ml, were also used to test the influence of BG concentration in the project. Solution B was 1-octanol saturated with water. 

\subsection{Extraction of acid from the flow and colorimetric reaction in droplets}
Once colored droplets were formed, acid extraction was from solution C , which had been prepared by dissolving a predetermined amount of various acids in 1-octanol. Solution C was injected into the fluid chamber at a constant flow rate of 50 ml/hr to replace solution B, as shown in Figure \ref{fx0}b. The acid concentration gradually increases at the mixing front of the solution B and C, due to the Tayler-Aris dispersion.\cite{taylor1953dispersion,aris1956dispersion}. When the injected solution C was in contact with the droplets on the substrate, acid extraction and its reaction with BG in the droplets were triggered. The acid concentration in solution C was constant due to the large volume of solution C compared to the small volume of the droplets. The color of blue droplets gradually changed to yellow as a result that the acid concentration in the droplet reached the critical concentration ($C^*$). The process of the droplet decoloration with different concentrations of acids was recorded using an upright optical microscope equipped with a high-resolution camera (Nikon H600l). The recorded images and videos were analyzed by ImageJ for further evaluation.

\subsection{Determination of temporal solute concentration in the flow}
A fluorescence dye, Rhodamine 6G (R6G), was used as the marker to trace the concentration of the solute in the flow at the mixing front of two solvents. R6G was directly dissolved in pure octanol as the tracer solution. The fluid chamber was firstly filled with pure octanol. Afterward, octanol was replaced by a flow of the tracer solution into the fluid at the flow rate of 50 ml/hr, the same as other experiments. The exposure time was 200 ms for each concentration. The fluorescence emission intensity of the tracer solution was detected and recorded by the microscope (Nikon H600l) equipped with a fluorescence filter. The recorded videos were analyzed by ImageJ. Three different concentrations of the R6G ($1\times 10^{−5}$ g/L, $5\times 10^{−5}$ g/L, and $1\times 10^{−4}$ g/L) were tested with 3 repeats for each concentration to verify the reproducibility of the results. The exposure time was 200 ms for each concentration. The fluorescence emission intensity of the tracer solution was detected and recorded by the microscope (Nikon H600l) equipped with a fluorescence filter.

\section{Results}

\subsection{Temporal concentration of the solute in the external flow}
\begin{figure}[htp]
	\includegraphics[trim={0cm 0cm 0cm 0cm}, clip, width=0.75\columnwidth]{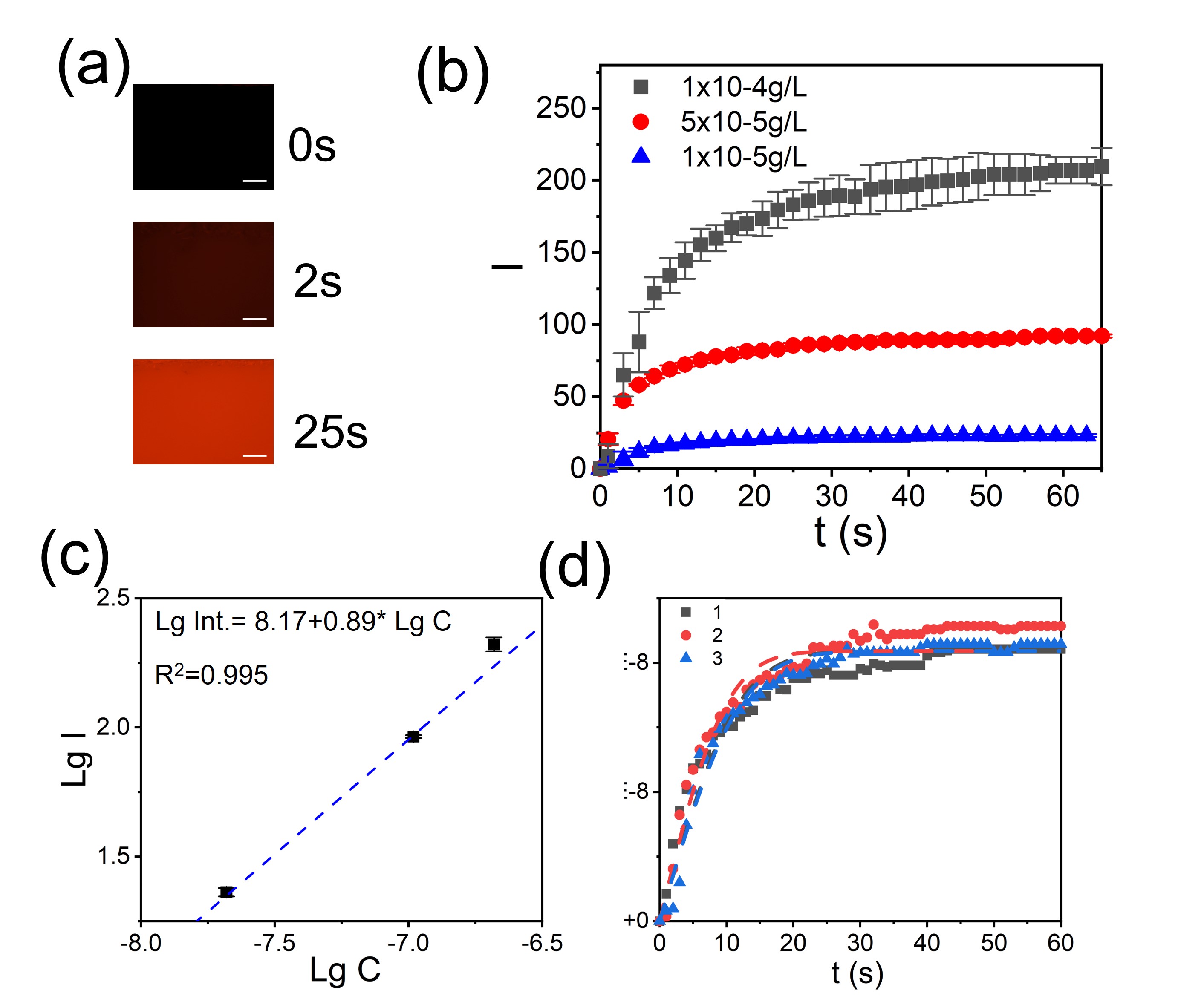}
	\caption{ (a) The images of the fluorescence intensity changes during displacing pure octanol by the dye solution. (b) The fluorescence intensity (I) from the solution as a function of time. The dye concentration was $1\times10^{−5}$, $5\times10^{−5}$, and $1\times10^{−4}g/L$. (c) Logarithmic plot of maximal fluorescence intensity with the concentration of the dye in the solution. (d) Error function fitted (dashed line) dye concentration as a function of time at the dye concentration of $1\times10^{−5}g/L$.}
	\label{SI}
\end{figure}

The temporal concentration of the analyte in a displacing flow supplied to surface nanodroplets is characterized by the fluorescence intensity of a dye solute in the solution.  The plots in Figure \ref{SI} show that at given flow rates and chamber dimensions in our experiments, the analyte concentration increased from 0 to the maximum after a transient period. Here the time of 0 is defined to be the moment when the fluorescence intensity appeared in the field of view. At the completion of the flow displacement, the final fluorescence intensity corresponds well to the concentration of the dye in the solution (Figure \ref{SI}(c)). \\
In the laminar flow regime, the temporal concentration of the analyte in the displacing flow is well-described by Tayler-Aris dispersion.\cite{dyett2018growth} The real-time concentration of the analyte as represented by the fluorescence intensity can be fitted with an error function. 
\begin{equation}
C_o(t)=C_{ini}\times Erf(\frac{t}{\tau})
\label{e5}
\end{equation}
Here, $C_o(t)$ is the real-time compound concentration in the external solution in contact with the surface nanodroplets at time $t$ (in second). $C_{ini}$ is the organic acid concentration in solution, and is also the concentration when $C_o(t)$ reaches the maximum. $\tau$ is a characteristic timescale determined by the flow condition. From the fitting shown in Figure \ref{SI} (d) and SI-1a and b, $\tau$ was determined to be around 12.87 seconds.\\ 
The temporal concentration of the model analytes of organic acids in our following experiments is expected to follow the same error functions for the dye concentration in the displacing flow. The acid concentration will increase initially from 0 to $C_{ini}$ in the oil solution and afterward remained constant at $C_{ini}$. The acid concentration $C_o(t)$ in the displacing flow will be determined by Eq.\ref{e5} with the same $\tau$ based on the dye concentration.

\subsection{Partition-dominated droplet decoloration at high concentration of acids}
\begin{figure}[htp]
	\includegraphics[trim={0cm 0cm 0cm 0cm}, clip, width=1\columnwidth]{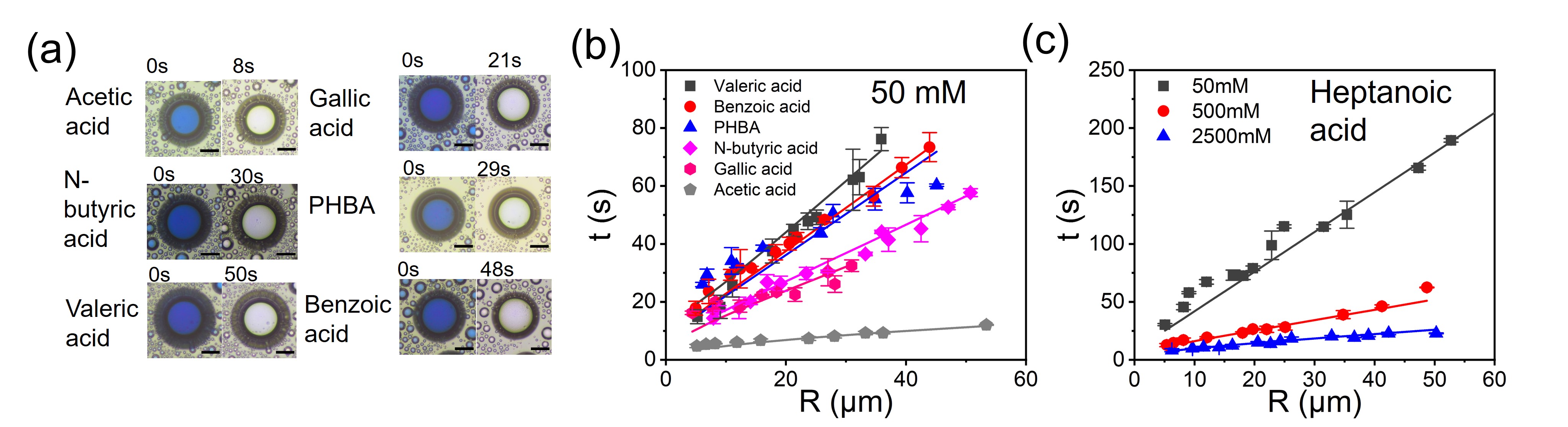}
	\caption{ (a) The optical images and the time scale of the droplets' decoloration in contact with the oil containing 50 mM of different acids. (b)  The plot of droplet decolor time as a function of droplet size after injecting oil containing 50 mM of acids. (c) The plot of droplet decolor time as a function of droplet size after injecting oil containing 50 mM, 500 mM and 2500 mM heptanoic acid. }
	\label{highconc}
\end{figure}

The organic acids used in our experiments can be defined as two groups, the alkyl acid group and the aromatic acid group. The molecule structure, partition coefficient (LgP in octanol and water) and the dissociation constant (pKa) of those acids in water are shown in Table \ref{t1}.\\ 
We first compared the decoloration time of the droplets at a given concentration of acids. A representative case in Figure \ref{highconc}a shows the color of droplets in contact with the flow of acid solution at $C_{o}$ of 50 mM. The droplets were 25 $\mu m$ in radius. All the droplets changed from blue to colorless within 100 s. Figure \ref{highconc}b shows the plot of the droplet decoloration time as a function of the droplet size. The droplet decoloration is systematically slightly faster from valeric to gallic acid. There is no significant difference in decoloration time for other types of acids, except acetic acid. We attribute the fast decoloration of acetic acid to its low partition coefficient in water (LgP \textless 0). At an even higher concentration (100 mM) of PHBA and benzoic acid, the decoloration time of the droplets is also close to each other, as shown in SI-2a.\\
Figure \ref{highconc}c shows the plot of the droplet decoloration time as a function of the droplet size for heptanoic acid at $C_o$ of 50 mM to 2.5 M. Here, 50 mM is the lowest concentration of heptanoic acid that can lead to decoloration, defined as the limit of detection (LoD). Compared to others acids, it takes the longest time for the droplets to decolor at 50 mM. The slowest decoloration may be due to the highest partition coefficient of heptanoic acid (least soluble in water) among all acids.

\begin{figure}[htp]
	\includegraphics[trim={0cm 0cm 0cm 0cm}, clip, width=1\columnwidth]{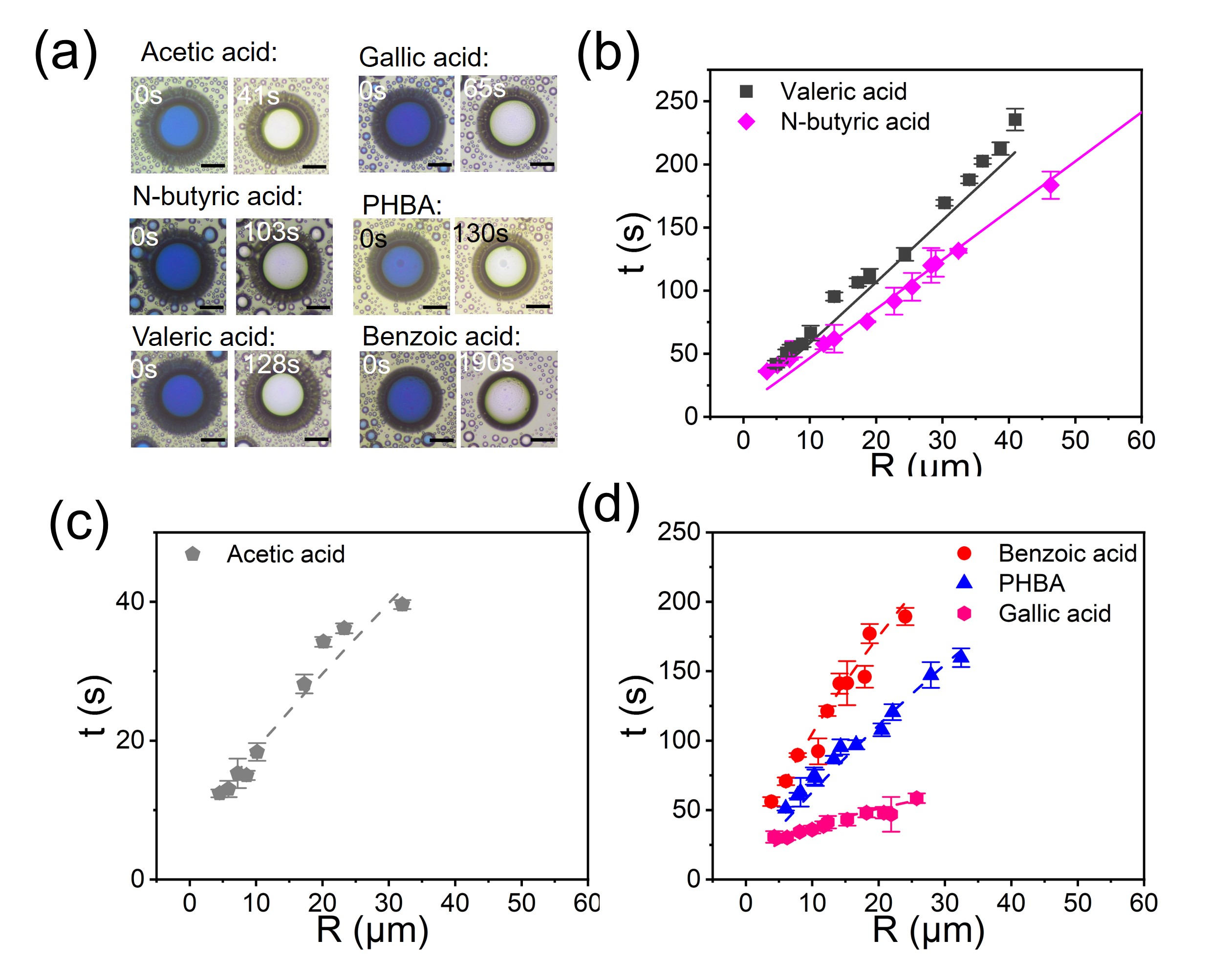}
	\caption{ The optical images of the same sized droplets (R=25 $\mu m$) that changes color at 5mM of acid in oil. a Time series images of the functional droplets at R=25 $\mu m$ change color after extraction of 5 mM aromatic acid. The time series images of the decolor time of the droplet in contact with oil containing different acids (b gallic, c PHBA and d benzoic acid) at 5 mM. Scale bar: 25 $\mu m$. (b) to (f) shows the individual acid fitted from Eq.\ref{e3} at given concentrations. }
	\label{f2}
\end{figure}

To determine the decoloration time, we need to consider the acid concentration in the droplet with time. In our experiments, the acid concentration was not constant in the initial 25 s. As the solution of acid in octanol solution was introduced into the chamber, the acid concentration changed from 0 to $C_o$ at the mixing front, where the extraction already happened. The concentration of acid ($HA$) in the oil solution $C_{o}$ is a function of $t$, determined by the flow rate of the acid solution. The flow rate was same in all of our experiments, giving the same profile of the acid concentration. Thus, we replace $C_{o}$ by $C_{o}(t)$ in Eq.\ref{e0}:
\begin{equation}
\frac{dC_w(t)}{dt}=\frac{3}{R}{\beta}(\frac{C_o(t)}{p}-C_w(t))
\label{e2}
\end{equation}
Here, $C_{o}(t)$ is represented in Eq.\ref{e5}. The extracted acid dissociates, producing protons that lead to the colorimetric reaction. The concentration of acid that is sufficient for droplet decoloration is defined as critical acid concentration $C^*_w$, calculated by pKa of a given acid and the pH value of the pH indicator.\cite{wei2020integrated} When pH value is 3.8, the proton concentration is $1.85 \times 10^{-4}$M, the droplet finished the decoloration. Then, by using the simple acid dissociation equation, $C^*_w$ can be obtained. The values of $C^*_w$ for all the acids are listed in Table \ref{tt1}. LgP and pKa of a given acid were found from the data available in the literature, as listed in Table \ref{t1}. $C_o$ is known as our experimental condition, Eq.\ref{e2} can be integrated to obtain the decoloration time $t_{decoloration}$ for $C_w(t)$ to increase from 0 to $C_w^*$. The interfacial transfer coefficient $\beta$ is a fitting parameter.\\
The solid lines in Figure \ref{highconc}b and c are fitted by the theoretical model analysis of Eq.\ref{e2}. The predicted $t_{decoloration}$ from Eq.\ref{e2} is in good agreement with the experiment results. The difference in decoloration time of different acids is attributed to the variation in the partition coefficient of the acid in water and octanol.\\
At the acid concentration from 50 mM and above, the solid lines shown in the plots (Figure \ref{f2}b, c and SI-2a) fit all well with the theoretical model analysis based on Eq.\ref{e2}. The results show that the dependence of the decoloration time on droplet size can be described well simply based on the mass transfer from the displacing flow into the droplets through the partition. Coupled with the temporal concentration profiles of the solute in the displacing flow, the established partition mechanism for microextraction of droplets is sufficient to predict the timescale required for droplet decoloration. Any possible effect of chemical reaction in the droplets can be neglected at a high concentration of acids. At a high concentration of the solute in the surrounding phase, the chemical reaction inside the droplets does not show a significant impact on the extraction rate of the acid.\\  

\subsection{Enhanced partition at low concentration of acids}
Interesting phenomena of enhanced extraction are discovered at low concentrations of acids. As shown in Table \ref{t1} and \ref{tt1} for 5 mM acid solution, the concentration of acids in water droplets ($C_w$) was calculated from the partition coefficient of the acids and the dissociation constant of the acids at the equilibrium state. $C^*_w$ was estimated from pKa of the indicator in water. At low acid concentration, the extracted acid in a droplet would not reach critical concentration $C^*_W$ if the only partition is responsible for the extraction. At the concentration of extracted acid in water ($C_w$) that was lower than the predicted critical concentration of the acid ($C^*_w$) for droplet decoloration, we found that the droplets can still lose color. For the least water-soluble acid (heptanoic acid), the decoloration at low acid concentration was found at 50 mM. The extracted acid was expected to be $2 \times 10^{-4}M$ according to the partition, lower than $C^*$($6 \times 10^{-4}M$). The results suggest that the pure partition is not sufficient to produce enough acids in the droplets for decoloration. An additional pathway must exist to provide more protons in the droplets for color change.  

\begin{table}[ht]
 \centering
 \caption{Decoloration time of 5 $\mu m$ droplets at 5 mM of acid and the acid concentration ($C_w$) in the droplets calculated from partition coefficent LgP.}
\begin{tabular}{p{2cm}|p{1.5cm}|p{1.5cm}|p{1.5cm}| p{1.5cm}| p{1.5cm}| p{1.5cm}}
\hline
Acid& Acetic & N-butyric & Valeric & Gallic & PHBA & Benzoic \\
\hline
Time (s)&12.7& 41& 42.8&30.7&48&63\\
\hline
$C_w$ (mM)&6.25& 0.81& 0.2&0.54&0.13&0.07\\
\hline
$C^*_w$ (mM) &1.44&1.66&0.6&0.7&0.9&0.4\\
\hline
\end{tabular}
\label{tt1}
\end{table}

Apart from the droplet decoloration at unexpectedly low concentrations of acids, reducing the concentration of acid in the displacing flow slows down droplet decoloration. For example, a 5-$\mu m$ droplet became colorless at 63 s from 5 mM benzoic acid, in contrast to the decolor time of 18 s from 50 mM acid in the flow. As $C_o(t)$ reached the plateau, it is expected that partition equilibrium can be reached in a few seconds at most due to the small dimension of the droplets. However, the decoloration time as listed in Table \ref{tt1} is much later than the moment when the acid concentration reaches equilibrium in the surrounding. 

\subsection{Interfacial partition enhanced by nanodroplet reaction }
\begin{figure}[htp]
	\includegraphics[trim={0cm 0cm 0cm 0cm}, clip, width=1\columnwidth]{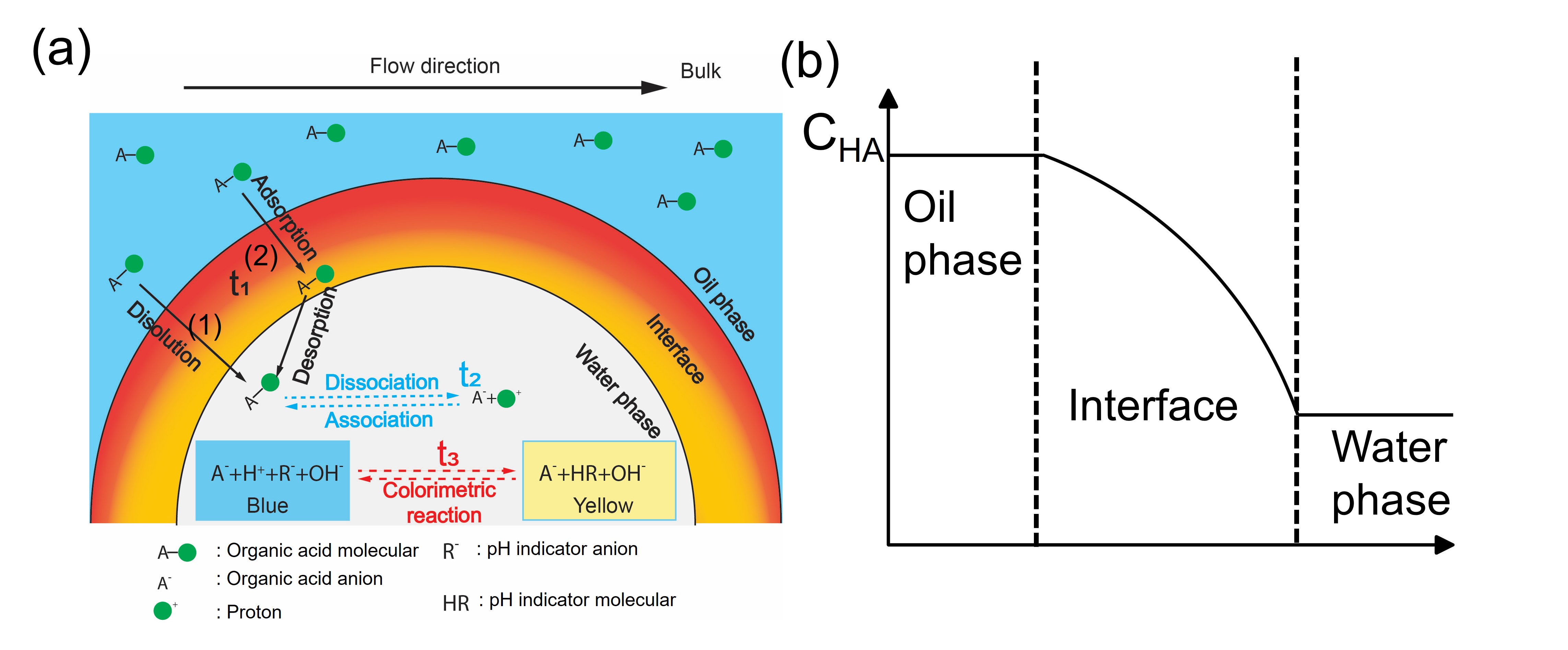}
	\caption{(a)Sketch of four steps in nanoextraction and colorimetric reaction in droplets. The acid molecules penetrated the oil-water interface by directly dissolving and surface adsorption-desorption (step 1: $t_1$). Then acid dissociated to release protons (step 2: $t_2$). The protons react with the BG, leading to a color change (step 3: $t_3$). The total time scale of the colorimetric reaction is the combination of those four processes ($1/t$=$1/t_1$+$1/t_2$+$1/t_3$). (b) The concentration gradient of the acid at the interface.}
	\label{sk2}
\end{figure}

To rationalize the extraction in droplets at low concentration of acids, we propose interfacial partition enhanced by nanodroplet reaction as below. The droplet decoloration process is decoupled into several steps as depicted in Figure \ref{sk2}.  
(1) Partition at the rate of $1/t_1$: A fraction of acid ($HA$) molecules dissolve into the aqueous droplets through partition. Those amphiphilic acid molecules may preferentially stay at the interface due to the interfacial adsorption. The mass transfer process can be expressed as:
\begin{equation}
\ce{$HA_o$ ->$HA$_w}
\label{partition}
\end{equation}
(2) Dissociation at the rate of $1/t_2$: The acids adsorbed at the interface or dissolved in the droplets dissociate and form $H^+$ and $A^-$. The dissociation rate is determined by the intrinsic dissociation constant($r_1$) of the acids at the droplet surface, which is a very fast. Here, the dissociation rate $1/t_2$ can be expressed as $\frac{d[H^+]}{dt}$. 
\begin{equation}
\ce{$HA_w$ ->[{r_1}]H^+ + A^-}
\label{er1}
\end{equation}
(3) Colorimetric reaction at the rate of $1/t_3$: In aqueous droplets, the pH indicator $R^-$ in the droplets combines with $H^+$ and converts to a product of HR. Here, the reaction rate $r_2$ can be expressed as $\frac{d[HR]}{dt}$.
\begin{equation}
\ce{H^+ + R^- (blue) ->[{r_2}]HR (yellow)}
\label{er2}
\end{equation}
The influence from $1/t_2$ and $1/t_3$ is not evident at the high concentration of the solute in the surrounding phase, because the time is short for the acid concentration of the solute in the droplets to reach a level for decoloration. However, $1/t_1$, $1/t_2$ and $1/t_3$ can not attribute to the rate-limiting step for decoloration at a low concentration of acid, because the decoloration is also much slower than the partition, as shown in Table \ref{tt1}. Once acid are extracted in water droplets, the dissociation and the protonization of the indicator can be completed on a timescale of less than 1 ms.

The proposed interfacial enrichment of acids may explain both the long duration of decoloration and the high sensitivity of droplet decoloration to the acid at low concentrations. We noticed that the long duration of decoloration occurs for acids with a positive partition coefficient which is more concentrated in the surrounding octanol than in water droplets. There is a boundary layer with a concentration gradient of acids that is lower in water droplets, and higher in the oil phase in the external solution, as sketched in Figure \ref{sk2}b. \\
The concentration inside the droplet is regulated by the acid dissociation and reaction between protons and the indicator. The colorimetric reaction consumes the extracted acid, driving a net influx of the acid into the droplet. Such influx is slow, because the transfer through the interface of the droplets remains quasi-equilibrium. Therefore, step 1 is the rate-limiting step.\\
The above partition shifting of acids to the droplets could happen, as the liquid-liquid interface is permeable and dynamic. In the absence of chemical reactions (step 2 and 3) in droplets, the amount of molecules into the droplet is same as that of molecules out of the droplets at the equilibrium of partition. However, for reacting droplets, the fast chemical reaction through step 2 and then 3 cleans up the captured molecules quickly from the interface, so the droplets surface is adsorbed and reset to accommodate more acid molecules from the outside solution. The droplet decoloration can take place even if the concentration of acid in bulk is below the level predicted by pure partition, based on the solubilities in two phases.\\
Although not required, the interfacial activity of the acids may also contribute to the enhanced microextraction of the droplets. Several acids are like surfactants, accumulating at the droplet interface. Those amphiphilic acid molecules may preferentially stay at the interface, due to the interfacial adsorption. The transfer through the oil-water interface of the droplets remains at adsorption-desorption equilibrium.

\subsection{Theoretical decoloration rate coupling interfacial partition and chemical reaction in droplets}
At low concentration of acids, the reaction rate may not be simply neglected for acids. As discussed above, the chemical reaction may drive the extraction of the acid to the droplets. 
$C_w(t)$ is expressed as:
\begin{equation}
\frac{dC_w(t)}{dt}=\frac{3}{R}\beta(\frac{C_o(t)}{p}-C_w(t))+rC_w(t)
\label{e3}
\end{equation}
The first term on the right determines acid transfer (partition and interfacial transfer). The second term determines reaction in the droplets (acid dissociation and pH indicator protonation).\\
A reaction rate $r$ of ($r$=$r_1 \times r_2$) accounts for the effects from acid dissociation and chemical reaction in the droplets. Here, $r_1$ represents the acid dissociation rate and $r_2$ represents the reaction rates of acid and the indicator in droplet. Both $r_1$ and $r_2$ defined in Eq.\ref{er1} and \ref{er2} are given by the acid dissociation and by the reactant in the droplets, respectively. Eq. \ref{e3} can be integrated to obtain the decoloration time $t_{decoloration}$, that is, the time for $C_w(t)$ to increase from 0 to $C_w^*$. The fitting parameters are the apparent partition coefficient $p^*$, the reaction rate r and mass transfer coefficient $\beta$. The shift of distribution coefficient from p to $p^*$ accounts for more acids extracted into the droplet to the level of $C^*_w$. \\

\subsection{Rate-limiting step in droplet decoloration at low acid concentration}
\begin{figure}[htp]
	\includegraphics[trim={0cm 0cm 0cm 0cm}, clip, width=1\columnwidth]{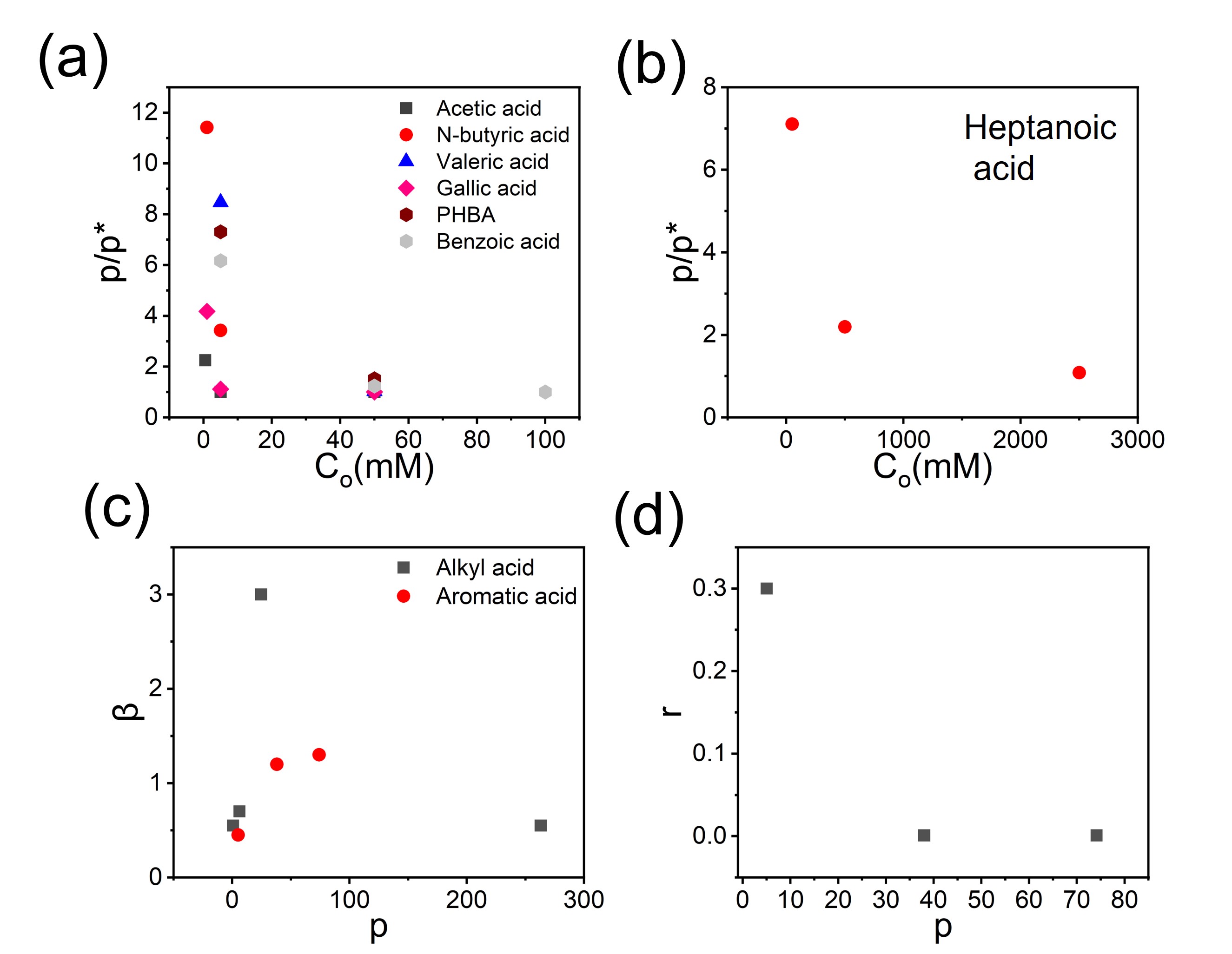}
	\caption{The distribution coefficient shift down with the decreasing of the acid concentration in oil ($C_o$) for heptanoic acid (b) and other acids(a), respectively. (c)The value of $\beta$ from the fitting of Eq.\ref{e2} at high concentration as a function of p for the alkyl and aromatic acids. (d) The value of r from the fitting of Eq.\ref{e3} at 5 mM as a function of apparent distribution coefficient ($p^*$) for the aromatic acids.}
	\label{f3}
\end{figure}

Eq.\ref{e3} show good agreement with our measured decoloration time for all types of acids, although decoloration time varies clearly with the properties of acids. As shown in Figure \ref{f2}a, the droplet decoloration becomes faster for more hydrophilic acid. For a given size of the droplet (R of 25 $\mu m$), the order of decoloration is acetic acid, gallic acid, n-butyric acid, PHBA, valeric acid and benzoic acids.\\
At the low acid concentration of 5 mM in Figure \ref{f2}(b-d), the decolor time from the two long-chain alkyl acids (n-butyric acid and valeric acid) can be described by Eq. \ref{e3}, as shown in Figure \ref{f3}c. Here, the value of the fitting parameter apparent partition $p^*$ is much lower than the intrinsic coefficient of the acid in octanol and water in the absence of chemical reactions. The other fitting parameter r is 0 from these two alkyl acids. Similar to surfactant, the hydrophilic part (carboxylic acid group) of these acids can penetrate the oil-water interface and stay in the water phase, while the hydrophobic component remains in the oil phase. The rates of acid dissociation (step 2) and colorimetric reaction (step 3) in the droplets are fast compared to interfacial transfer. So the second term on the left side of Eq. \ref{e3} can be neglected (r=0).\\
As for acetic acid, its solubility in water is much higher than in the surrounding. Compared to water-less soluble valeric acid and butyric acid, aromatic acids have the less surface activities. Droplet decoloration from acetic acid and aromatic acid show good agreement with the prediction from Eq. \ref{e3}. Different from r=0 for valeric acid and butyric acid, a positive value of r was obtained in Eq. \ref{e3}, as shown in Figure \ref{f3}d. Park et al. demonstrated that the dissociation of acetic acid is a very fast process, completed in few picoseconds.\cite{park2006dissociation} However, the effect from the reaction rate r on the decoloration time can be revealed in decoloration rate, possibly due to the fast partition process of acetic acid and aromatic acids in droplets. We also found that r decreases with $C_o$, as shown in Table \ref{tp1} and \ref{tp2}, suggesting less influence from chemical reaction in droplets and more solubility-dominated partition at higher $C_o$.\\
The partition shifting happens to the droplets at low concentrations specific of acids. The apparent distribution $p^*$ obtained at all concentrations, is shown in Figure \ref{f3} a and b. The revers of $p^*$ represents the concentration in droplets over that in the surrounding. As the acid concentration decreases, the partition shifting (p/$p^*$) becomes more significant. Up to 11 times larger $p^*$ was found for butyric acid at $C_o$= 1 mM in Figure \ref{f3}. p and $p^*$ become the same at high concentrations of acids, for example, at 2.5 M for heptanoic, and 100 mM for benzoic acid and PHBA, 50 mM for other acids.

\begin{table}[]
\centering
\caption{The mass transfer coefficient ($\beta$), reaction coefficient (r) and apparent distribution constant ($p^*$) of acetic acid and aromatic acid at the concentration ($C_o$) in the octanol solution.}
\begin{tabular}{P{2cm}|P{2cm}|P{2.2cm}|P{2.2cm}|P{2.2cm}|P{2.5cm}}
\hline
$C_o$ (mM)            & Parameters & Acetic acid &Gallic acid & PHBA  & Benzoic acid \\
\hline
\multirow{3}{*}{0.5} & $\beta$ & 0.55       &         &      &           \\
                     & r       & 0.005      &            &      &             \\
                     & $p^*$       & 0.3         &            &      &             \\
                     \hline
\multirow{3}{*}{1}   & $\beta$ &           & 0.45        &      &            \\
                     & r       &            & 0.01        &      &            \\
                     & $p^*$       &            & 1.2         &      &             \\
                     \hline
\multirow{3}{*}{5}   & $\beta$ & 0.55        & 0.45        & 1.2   & 1.3          \\
                     & r       &    0.01         & 0.03         & 0.001 & 0.001        \\
                     & $p^*$       & 0.67        & 4.5         & 5.2   & 12           \\
                     \hline
\multirow{3}{*}{50}  & $\beta$ & 0.55        & 0.45        & 1.2   & 1.3          \\
                     & r       &           &            &      &             \\
                     & $p^*$       & 0.67        & 5.01        & 25    & 60           \\
                     \hline
\multirow{3}{*}{100} & $\beta$ &           &            & 1.2   & 1.3          \\
                     & r       &            &            &      &             \\
                     & $p^*$       &            &            & 38    & 74   \\
                     \hline
\end{tabular}
\label{tp1}
\end{table}
\begin{table}[]
\centering
\caption{The mass transfer coefficient ($\beta$), reaction coefficient (r) and app rent distribution constant ($p^*$) of long-chain alkyl acid at the concentration ($C_o$) in the octanol solution.}
\begin{tabular}{P{2cm}|P{2.2cm}|P{3cm}|P{2.2cm}|P{3cm}}
\hline
$C_o$ (mM)            & Parameters  & N-butyric acid & Valeric acid & Heptanoic acid \\
\hline
\multirow{2}{*}{1}   & $\beta$  & 0.7            &             &         \\
                
                     & $p^*$           & 0.54           &             & \\
                     \hline
\multirow{2}{*}{5}   & $\beta$     & 0.7            & 3           & \\
                    
                     & $p^*$         & 1.8            & 2.9          &\\
                     \hline
\multirow{2}{*}{50}  & $\beta$      & 0.7            & 3            &0.55 \\
                    
                     & $p^*$             & 6.2            & 24.5         &37  \\
                     \hline
\multirow{2}{*}{500} & $\beta$             &              &             &0.55 \\
                     
                     & $p^*$                  &               &            & 120 \\
                           \hline

\multirow{2}{*}{2500} & $\beta$        &              &             &0.55\\
                    
                     & $p^*$                   &               &            & 243  \\
                     \hline                                    
\end{tabular}
\label{tp2}
\end{table}

\subsection{Acid specificity of mass transfer coefficient $\beta$}
The mass transfer coefficient ($\beta$) shown in Figure \ref{f3}c mainly reflects the rate of the acid crossing the interface. We found that $\beta$ is independent of the concentration in the oil phase ($C_o$) for a given acid. But for different acids, expect heptanoic acid, $\beta$ increases with the distribution coefficient p. As shown in Figure \ref{f3}c, possibly acids with a lower distribution coefficient (p) established as lower the acid concentration gradient between the droplet and the surrounding for a slow transfer of the acid. The interfacial activity of the acid also influences the transfer coefficient $\beta$. At the same p, the coefficient $\beta$ of aromatic acids with lower interfacial activity is lower than that of alkyl acids. On the other hand, the long hydrophobic chain in the heptanoic acid molecule makes it much more likely to stay on the oil phase than in water droplets. Hence, transferring heptanoic acid into water droplets becomes energetically unfavorable, leading to a very low $\beta$.

\subsection{Effects of reactant concentration in droplets on reaction rates}

\begin{figure}[htp]
	\includegraphics[trim={0cm 0cm 0cm 0cm}, clip, width=1\columnwidth]{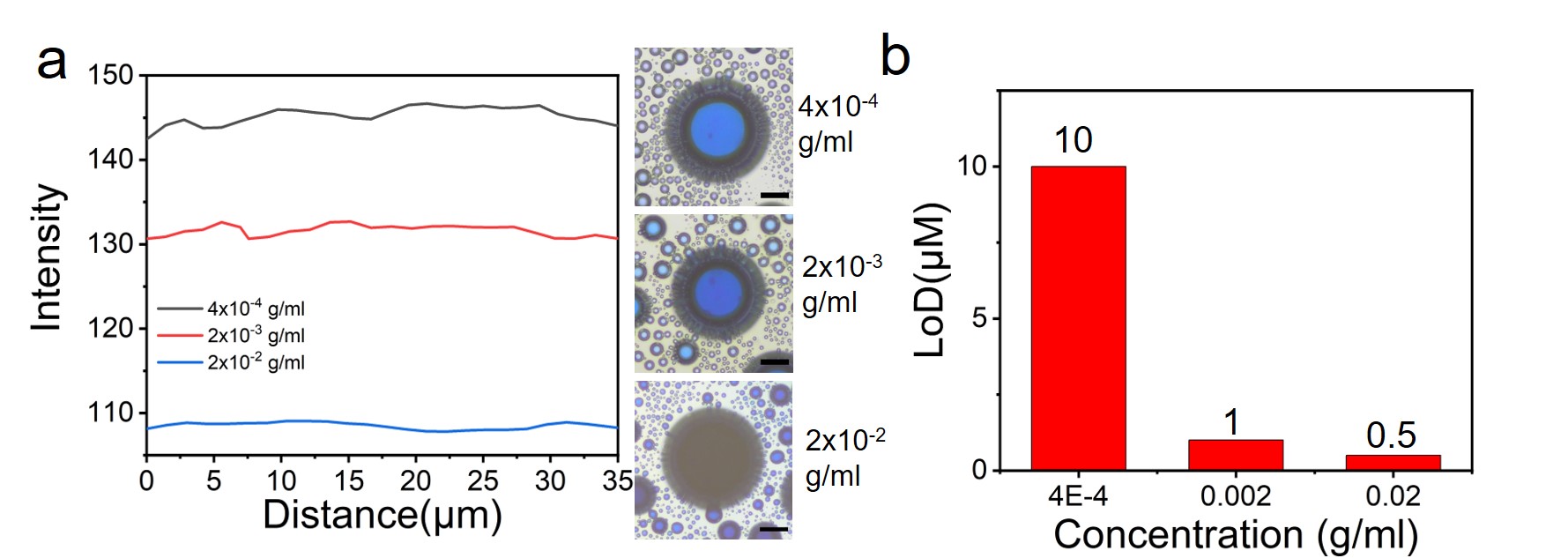}
	\caption{The initial color intensity along droplet diameter and the images (a) and the limit of detection (LoD) (b) for benzoic acid of the droplets that formed by solution A containing BG at the concentration of 0.02g/ml, 0.002 g/ml and 0.0004 g/ml. Scale bar: 25 $\mu m$. Here, distance 0 means the edge of the droplets.}
	\label{fx00}
\end{figure}

In the above, the concentration of the BG in the droplets was the same in all of the experiments (0.02 g/ml in solution A). Considering the crucial role of chemical reactions in the droplets for acid detection, another important parameter to influence the detection sensitivity is the concentration of BG in the droplets. We varied the pH indicator concentrations by changing the concentration of BG in solution A. After the formation of the droplets by solvent exchange, a higher concentration of BG showed a darker color (lower color intensity), as shown in Figure \ref{fx00}a. We used benzoic acid as the model analyte to measure the LoD under different BG concentrations in solution A. The results in Figure \ref{fx00}b demonstrates that a higher concentration of BG had a higher LoD.\\

Higher sensitivity (lower LoD) in a lower concentration of BG in the droplets is attributed to a shorter time to reach the critical concentration. The decoloration of the droplets is determined by the ratio of the concentration of $HR$ and $R^-$ ($\frac{[HR]}{[R^-]}$), as shown in the Eq.\ref{er2}. As the [$R^-$] increased when the BG concentration increased in the droplets, the amount of the $HA$ for droplets to accomplish the decoloration is also increased.\\
More HA molecules need a longer time to reach the critical concentration. When the concentration of the HA increases, the reaction rate ($r_2$) also increases. However, compared to the reaction rate, the acid dissociation rate ($r_1$) is faster.\cite{park2006dissociation} Considered the organic acids used in this work are weak acid and thus the acid dissociation is partial, more acid molecules accumulated in the droplet. Those redundant HA molecules slow down the mass transfer rate due to the decreases of the driven force ($C_o$-$C_w$).

\section{Conclusions}

This work experimentally and theoretically investigates the rate and efficiency of integrated extraction and detection of analytes in surface nanodroplets. Seven acids were used as the model analysts, and the extraction of the acid was coupled with a colorimetric reaction in surface nanodroplets. We found a significant shift in the partition of the acids into the reacting droplets at a low concentration of acid in the external solution.  The interfacial partitioning of the acids was proposed to be the rate-limiting step in integrated extraction and detection, and meanwhile contribute to the highly sensitive detection from the shifted partition of the acids.  The proposed theoretical model for the extraction process that takes the partition shift into consideration is in good agreement with experimental results. This work provides a comprehensive analysis for the enhance of extraction and sensitive colorimetric etection in nanodroplets.  
\section{Acknowledge}
The project is supported by the Natural Science and Engineering Research Council of Canada (NSERC) and Future Energy Systems (Canada First Research Excellence Fund). We thank Nikoo Moradpour for improving the language and Qiuyun Lu, Zhengxin Li, Jia Meng and Romain Billet for valuable discussion on the theoretical analysis.
\section{List of symbols}
\begin{table}[]
 \centering
\begin{tabular}{p{2.5cm}|p{10cm}}
Symbol & descrition \\
\hline
$\beta$& Mass transfer coefficent\\
$C^*$& Critical concentration of the acid in water\\
$C_o$& Acid concentration in oil phase\\
$C_w$& Acid concentration in water phase\\
$C_o(t)$& Acid concentration in oil phase as a function of time\\
$C_w(t)$& Acid concentration in water phase as a function of time\\
HA& Acid molecule\\
$H^+$& Proton\\
HR& Halochromic compound molecule\\
I& Fluorescence intensity of the dye solution\\
LgP& Partition coefficient, logarithm of p\\
LoD& Limit of detection\\
p & Distribution coefficient, $p=\frac{C_o}{C_w}$\\
$p^*$ & Apparent distribution coefficient\\
pKa& Dissociation constant\\
R& Lateral radius of the surface droplets\\
$R^-$& Anion of halochromic compound\\
r& Total reaction rate and in droplet\\
$r_1$& Dissociation rate of acid\\
$r_2$& Reaction rate of protons and $R^-$\\
t& Droplet decolor time\\
\end{tabular}
\label{tt}
\end{table}
\newpage
\begin{figure}[htp]
	\includegraphics[trim={0cm 0cm 0cm 0cm}, clip, width=1\columnwidth]{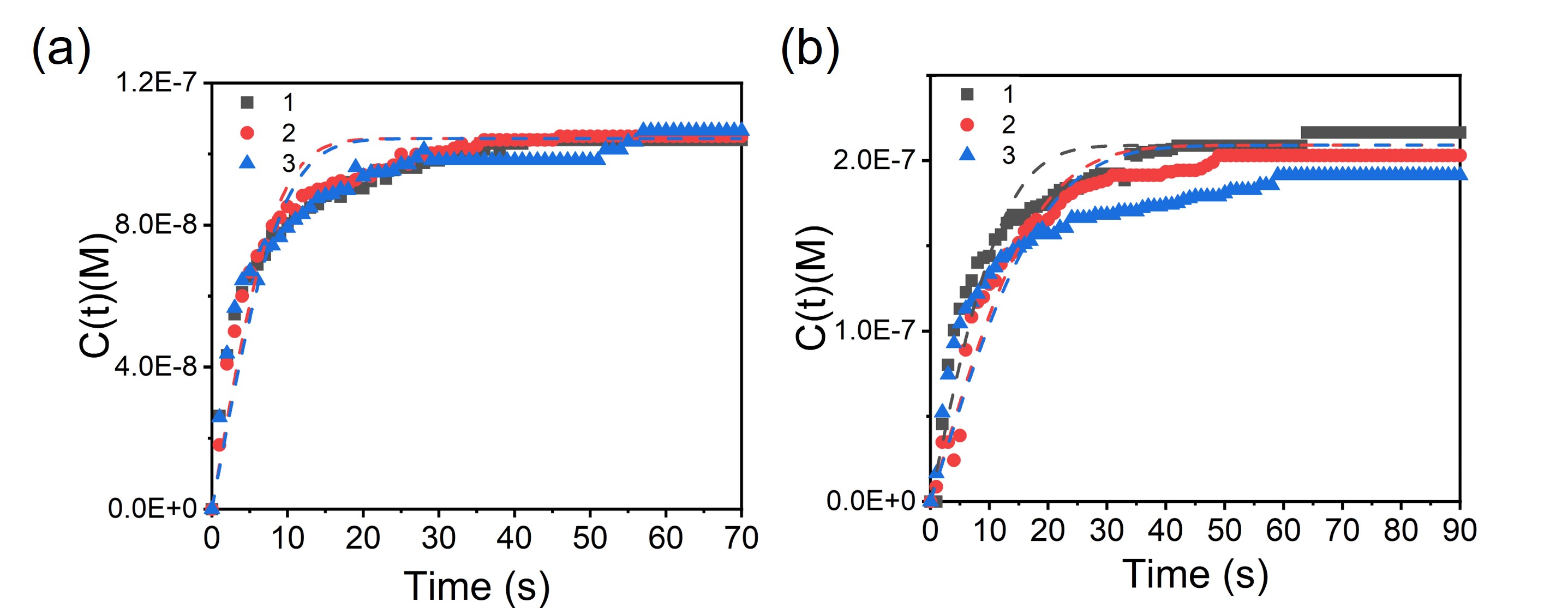}
	\caption{ (Error function fitted (dotted line) the fluorescence intensity increase from time 0 at dye concentration of $5\times 10^{-5}$ g/L (a) and $1\times 10^{-4}$ g/L (b).}
	\label{SI-1}
\end{figure}
\begin{figure}[htp]
	\includegraphics[trim={0cm 0cm 0cm 0cm}, clip, width=1\columnwidth]{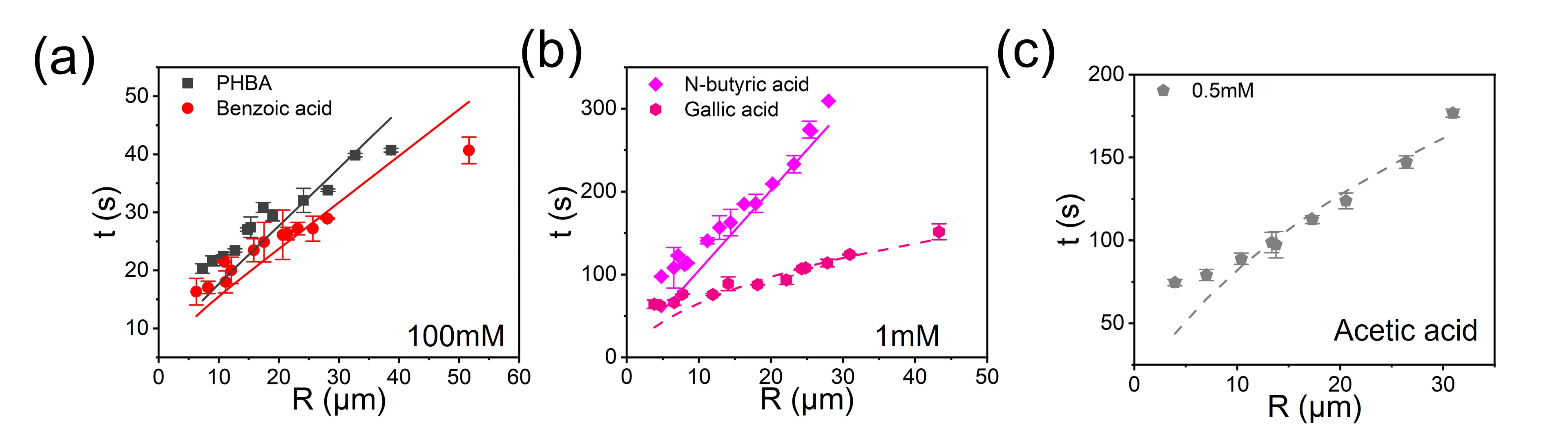}
	\caption{ The plot of droplet decolor time as a function of droplet size after injecting oil containing 100 mM of PHBA and benzoic acid (a), 1 mM of n-butyric acid and gallic acid (b) and 0.5 mM of acetic acid (c). The solid line is fitted from Eq.3. The dash line is fitted from Eq.7. }
	\label{SI-2}
\end{figure}
\bibliography{Literature}
\end{document}